\def\year{2021}\relax
\documentclass[letterpaper]{article} 

\usepackage{aaai21}  
\usepackage{times}  
\usepackage{helvet} 
\usepackage{courier}  
\usepackage[hyphens]{url}  
\usepackage{graphicx} 
\usepackage{natbib}  
\usepackage{caption} 
\title{A Unified Graph-Based Approach to Disinformation Detection using Contextual and Semantic Relations}

\author {
    Marius Paraschiv,\textsuperscript{\rm 1}
    Nikos Salamanos, \textsuperscript{\rm 2}
    Costas Iordanou, \textsuperscript{\rm 2} 
    Nikolaos Laoutaris,\textsuperscript{\rm 1}\\ 
    Michael Sirivianos \textsuperscript{\rm 2} \\
}
\affiliations {
    \textsuperscript{\rm 1} IMDEA Networks Institute, Spain 
    \textsuperscript{\rm 2} Cyprus University of Technology \\
    marius.paraschiv@imdea.org, nik.salaman@cut.ac.cy, costas.iordanou@eecei.cut.ac.cy,\\ nikolaos.laoutaris@imdea.org, michael.sirivianos@cut.ac.cy
}

\usepackage{graphicx}
\usepackage{subfigure}
\usepackage{booktabs} 

\usepackage{makecell}
\usepackage{amsmath}

\usepackage{amssymb}
\usepackage{arydshln}
\usepackage{breakurl}
\usepackage{amsthm}
\usepackage{xcolor} 

\theoremstyle{definition}
\newtheorem{definition}{Definition}[section]

\copyrighttext{}
\begin{document}

\maketitle

\begin{abstract}
As recent events have demonstrated, disinformation spread through social networks can have dire political, economic and social consequences. Detecting disinformation must inevitably rely on the structure of the network, on users particularities and on event occurrence patterns. We present a graph data structure, which we denote as a \textit{meta-graph}, that combines underlying users' relational event information, as well as semantic and topical modeling. We detail the construction of an example meta-graph using Twitter data covering the 2016 US election campaign and then compare the detection of disinformation at cascade level, using well-known graph neural network algorithms, to the same algorithms applied on the meta-graph nodes. The comparison shows a consistent 3\%-4\% improvement in accuracy when using the meta-graph, over all considered algorithms, compared to basic cascade classification, and a further 1\% increase when topic modeling and sentiment analysis are considered. We carry out the same experiment on two other datasets, HealthRelease and HealthStory, part of the FakeHealth dataset repository, with consistent results. Finally, we discuss further advantages of our approach, such as the ability to augment the graph structure using external data sources, the ease with which multiple meta-graphs can be combined as well as a comparison of our method to other graph-based disinformation detection frameworks.
\end{abstract}

\section{Introduction}
\label{introduction}

Social media have become a primary medium of interaction in modern society -- having started as a place for casual discussion and exchange of ideas. Their massive outreach and adaptability to individual users' preferences have made social networks indispensable for a wide range of political and activist groups, companies, governments, and mainstream news outlets. The social importance of these networks stems from the fact that they are more than just a space for public discussion. Social networks such as Twitter and Facebook have become the primary source of information on a global scale ~\cite{social_media1}. However, the spread of disinformation can have a severe negative social \cite{social_impact}, political \cite{political_impact}, and economic impact \cite{economic_impact1, economic_impact2}, thus it is currently treated as a fast-growing cyberthreat \cite{cyberthreat}. 

Current disinformation detection methods on social media primarily rely on using part of the available information either regarding users or local network structure (see Section \ref{related_work}), in order to perform the required analysis. Treating such features in isolation may lead to poor results as the manifestation of disinformation is characterized by a plethora of factors, both internal (user characteristics) and external (network characteristics and tweet content). 

In this paper, we set out to improve our capacity to detect disinformation in social media by proposing a unified methodology for both internal and external factors. To demonstrate the value of the proposed method, we focus on Twitter content. The primary action on Twitter is message posting by registered users (\textit{tweets}) or message sharing by others within the platform (\textit{retweets}). As such, posting information on a particular topic results in a \textit{root--tweet} (the original tweet that has been retweeted), followed by a series of retweets by other users. This series of retweet events forms a tree--graph structure which is known as \textit{retweet cascade}. Moreover, we construct and analyse a very large dataset related to 2016 US Presidential Election, consisting of 152.5M tweets by 9.9M users; In addition, we evaluate our method on the HealthRelease and HealthStory datasets, part of the FakeHealth repository \cite{FakeHealth}. 

In summary, we propose a graph data structure, named \textit{meta-graph} having the retweet cascades as nodes. Each cascade disseminates a root--tweet together with the web and/or media URL(s) embedded in that tweet. The node features contain: (i) information about the cascade graph structure, obtained by applying a graph embedding algorithm to each individual cascade; (ii) relational user information, extracted from the social network. The edges of the meta-graph represent relationships between cascades, such as structural similarity, number of common users (possibly indicating that the cascades originate from the same community), or content similarity (URL or text of the root tweet). One of the advantages of this approach is that it can be easily expanded with additional information from other datasets. Finally, in the US Elections dataset, many URLs corresponding to the retweet cascades that have been manually labeled as \textit{fake} or \textit{non-fake}, by several external fact-checkers ~\cite{MBFC, PolitiFacts, FactCheck, Snopes}, resulting in a total of 43,989 labeled data points (see Table \ref{tab:labellingstats}).

The construction of the meta-graph poses a series of non trivial challenges, such as the choice of content-embedding and graph-embedding algorithms, the selection of relevant features from the raw data or the filtering of initial edges -- all of which are described in detail in Section \ref{metagraph}. By using the meta-graph approach, we transform the cascade classification problem into a node classification task. In case retweet cascades are treated as independent graphs, then we are dealing with a graph classification task. Data points are represented by individual cascades, labeled by their content. The similarity between cascades is thus a learned feature, inherent to the model itself. There are however advantages to making these relations explicit, such as biasing the algorithm in the desired direction, as well as the capacity to provide additional, external, information to the meta-graph edges, otherwise unavailable to a classification algorithm. 

In Section \ref{experiments} we show a comparison between (i) classifying cascades in isolation (graph classification); (ii) classifying cascades as nodes of the mentioned meta-graph. We demonstrate that the additional relational information contained in the meta-graph leads to consistently higher classification results. For the graph and node classification tasks, we apply four well-known graph neural network algorithms, as they currently represent the state of the art in terms of graph analysis and prediction. For an overview of graph neural networks, we refer the reader to some of the recent review papers, such as \cite{gnn_review1, gnn_review2, gnn_review3}. As an additional step, to the initial cascade or meta-graph features we append the sentiment analysis scores, as well as the predicted sentiment. This is done so as to avoid instances where a cascade is classified as disinformation, even though the root tweet is highlighting the possible issues with a piece of information, showing clear disagreement. A final set of features comes from topic modeling, where we detect topics, among all tweet texts in the dataset, and assign their corresponding identifier to each cascade or meta-graph node. This provides additional information to the classification algorithm, and we show that this approach offers 1\% consistent improvements.

\noindent\textbf{Data availability:} Part of the US Elections dataset is publicly available~\cite{US2016Election} under proper restrictions for compliance with Twitter's Terms of Service (Tos), General Data Protection Regulation (GDPR). The FakeHealth repository is publicly available~\cite{FakeHealth}.

\section{Related Work}
\label{related_work}

\begin{table}[t]
\begin{center}
\resizebox{\columnwidth}{!}{%
\begin{tabular}{lccc}
\toprule
Work     & \multicolumn{1}{c}{\begin{tabular}[c]{@{}c@{}}User feature-based  \\ Detection\end{tabular}} & \multicolumn{1}{c}{\begin{tabular}[c]{@{}c@{}}Network-based\\ Detection\end{tabular}} & \multicolumn{1}{c}{\begin{tabular}[c]{@{}c@{}}Content-based\\ Detection\end{tabular}} \\
\midrule
\cite{ngram}             &                          &                           & {\color{blue}\checkmark}\\ \hdashline
\cite{text_style}        &                          &                           & {\color{blue}\checkmark}\\ \hdashline
\cite{lecunn}            &                          &                           & {\color{blue}\checkmark}\\ \hdashline
\cite{fakebert}          &                          &                           & {\color{blue}\checkmark}\\ \hdashline
\cite{castillo}          & {\color{blue}\checkmark} &                           & {\color{blue}\checkmark}\\ \hdashline
\cite{shu}               & {\color{blue}\checkmark} &                           &                         \\ \hdashline
\cite{user_based1}       & {\color{blue}\checkmark} &                           &                         \\ \hdashline
\cite{user_based2}       & {\color{blue}\checkmark} &                           &                         \\ \hdashline
\cite{user_based3}       & {\color{blue}\checkmark} &                           &                         \\ \hdashline
\cite{event_credibility} & {\color{blue}\checkmark} &  {\color{blue}\checkmark} &                         \\ \hdashline
\cite{subnet}            &                          &  {\color{blue}\checkmark} & {\color{blue}\checkmark}\\ \hdashline
\cite{tracing}           &                          &  {\color{blue}\checkmark} &                         \\ \hdashline
\cite{similar_to_us}     & {\color{blue}\checkmark} &  {\color{blue}\checkmark} &                         \\ \hdashline
\cite{hierarchical}      &                          &  {\color{blue}\checkmark} & {\color{blue}\checkmark}\\ \midrule
\textbf{This Work}       & {\color{blue}\checkmark} &  {\color{blue}\checkmark} & {\color{blue}\checkmark}\\ 
\bottomrule
\end{tabular}
}
\end{center}
\caption{Summary of the related work based on different methodologies that they utilize/combine to identify fake news in social media.} 
\label{tab:related_work}
\end{table}

Substantial effort has been put into studying false information detection in social media. A first set of approaches concerns performing feature-based detection at user level. \cite{ngram} apply various feature extraction methods to the content of social media posts to detect stylistic characteristics that may give away posts containing disinformation. While this is a significant intermediary step, it does not consider relational information between sources and platform users (in the case of Twitter). \cite{text_style} use a similar approach to split the problem into feature extraction and classification steps, while \cite{lecunn} use character-level convolutional networks that combine the two. A more recent approach is that by \cite{fakebert}, which uses the BERT language model~\cite{bert} to perform feature extraction. A detailed overview of similar methods can be found in~\cite{fakenews_review}. \cite{castillo} assess the credibility of posts based on the past behavior of users, the tweet content, and other external sources. \cite{shu} extract features based on particular users' profiles and determine which type of profile is more inclined to share false information. This is similar to the methods in~\cite{user_based1, user_based2, user_based3} for exploiting user-type characteristics to classify posted messages as false or not.

Social context-based methods deal both with relations between users sharing news as well as with information related to the user and the news itself (for example, the number of similar postings made by the said user in the past). \cite{event_credibility} propose studying the problem from a credibility perspective by performing credibility propagation along a network that comprises events, tweets, and users. The events in this case roughly correspond to the root tweet in our approach. However, their approach does not consider the rich structural information of the cascades as well as cascade similarity. \cite{subnet} propose a sentence-comment co-attention sub-network, aiming to explain why a piece of news has been detected by an algorithm as false. At the same time, \cite{tracing} take the approach of detecting false information based on its propagation patterns through the network. 

A similar method to the one discussed in this paper is presented by \cite{similar_to_us}. The authors use an embedding framework dubbed TriFN, modelling publisher-news relations and users' interaction with the particular pieces of news simultaneously. Compared to our proposal, it misses relational information between events (cascades), which can prove essential in assessing a user's past behavior, as well as the semantic features added by topic modeling and sentiment analysis. Another interesting approach is presented in \cite{hierarchical} where the authors build a hierarchical propagation network for false information and perform a comparative analysis between false and real news based on linguistic, structural and temporal perspectives.

The methods outlined above roughly fall into three categories: a) \textit{user feature-based detection} in which the aim is to use information at the user level (be it contextual or behavioral) to assess the type of content a user publishes; b) \textit{network-based detection} in which the relational context is used for content classification; and c) and \textit{tweet content-based detection}, in which linguistic features are exploited for performing classification at the text level. Our proposed meta-graph data structure provides an effective means to combine all three methodologies, fully utilizing content, context, and source information simultaneously, as depicted in Table~\ref{tab:related_work}.

\begin{table}[t]
	\begin{center}
				\resizebox{\columnwidth}{!}{%
					\begin{tabular}{ll|ll}
						\toprule						
						Root Tweets & 46,409        &  Root Users  & 8204    \\
						Retweets    & 19,588,072    &  Retweeters  & 3,630,992 \\						
						URLs (web and media)        & \multicolumn{3}{l}{43,989}  \\    
						\bottomrule
					\end{tabular}
				}
	\end{center}
	\caption{US Elections dataset: Retweet cascades with at least 100 unique retweeters}	
	\label{tab:retweets}
\end{table}

\begin{table}[t]
\begin{center}
\begin{tabular}{lrr}
\toprule
         & HealthRelease & HealthStory \\
\midrule
tweets   &   43,245         &  357,851     \\ 
replies    &   1418         & 23,632     \\ 
retweets   &   15,343         &   105,712   \\ 
\midrule
Total      &     60,006       &    487,195        \\
\bottomrule
\end{tabular}
\end{center}
\caption{FakeHealth repository: Tweets, replies and retweets in the collected datasets} 
\label{tab:fakehealth_data}
\end{table}

\section{Social Media Data Structure}
\label{social_media_data}

\subsection{US Elections dataset}
\label{subsec:twitter_data}

To analyze disinformation on Twitter, we collected a large number of tweets related to the 2016 US presidential election. During that period, state--sponsored disinformation campaigns are believed to have operated by spreading millions of tweets with ambiguous political content. Hence, Twitter account activity from that period provides us with valuable information for the analysis of disinformation spread in social media. The belief that many of the tweets from that period were spreading fake news has been furthered validated by the fact that Twitter has permanently deleted lots of them as part of its continuous efforts against disinformation and malicious activities on the platform\footnote{\url{https://about.twitter.com/en/our-priorities/civic-integrity}}. \\  

\noindent\textbf{Crawling:} In the period up to the 2016 US presidential election -- from September $21^{\text{st}}$ to November $7^{\text{th}}$, 2016 (with the exception of October $2^{\text{nd}}$ 2016) we collected 152.5M tweets (by 9.9M users) using the Tweepy\footnote{\url{https://www.tweepy.org/}} Python library for accessing the Twitter streaming API. To consider tweets to be politically related, they need to include words from a list of $77$ track terms used for the crawling. Track terms as ``hillary2016'', ``clinton2016'', ``trump2016'', ``election2016'' etc. would, with a large degree of confidence, return tweets that were indeed part of the intense political polarization and debate that took place during the election period. For each tweet, we stored 27 features related to a tweet (tweet--ID, tweet--text etc.), and to the Twitter account (user--ID, user screen name, number of followers, etc.) who posted that tweet. Moreover, we collected the ``Entities'' section, which contains several metadata such as the ``mentions'' and the URLs embedded in the text.

We note that, in this dataset, we have already identified 35.5K tweets from 822 state--sponsored Twitter accounts based on ground--truth data provided by Twitter itself -- a large sample of state--sponsored disinformation campaigns from ``troll'' accounts which operated during that period. Let us note that ``troll'' is any account that deliberately spreads disinformation, tries to inflict conflict or causes extreme emotional reactions. Hence, our dataset contains valuable information of ground--truth malicious activities which in fact were the subject of a previous study regarding the trolls' activities during the 2016 US election~\cite{US2016Election}. 

In the retweeting process, there are two actors; (i) the \textit{retweeter}; (ii) the \textit{root--user}, i.e., the user who posted the original tweet (\textit{root--tweet}). We concentrate our attention on retweets that are sufficiently rich in terms of the information transmitted and the population of users that acted as retweeters. For this reason, we concentrate on cascades for which the root--tweet contains at least one web or media URL. We also dropped cascades with less than 100 retweeters. Tweets that very few users have retweeted do not provide enough information, even collectively, regarding their political positions. Following this approach, we analyse 46.4K retweet cascades consisting of 19.6M tweets (Table~\ref{tab:retweets}). 

\subsubsection{Labeling URLs}
\label{labeling}
To perform supervised/semi-supervised learning, we need to label the collected URLs as \textit{``fake''}, \textit{``non-fake''} or \textit{``unknown''}. Towards that end, we apply the following methodology:

\begin{table*}[t]
\begin{center}
\begin{tabular}{lrrr}
\toprule
Label     & Initial Labeling & Manual Labeling & Difference \\
\midrule
Fake      & 4,386            & 4,556    & +170   \\ 
Non\_Fake & 915              & 1,969    & +1,054 \\ 
Unknown   & 38,688           & 37,464   & -1,224 \\ 
\midrule
Total     & 43,989           & 43,989   &        \\
\bottomrule
\end{tabular}
\end{center}
\caption{US Elections dataset: The number of labeled URLs obtained at each step of the labeling methodology.} 
\label{tab:labellingstats}
\end{table*}

\begin{table}[t]
\begin{center}
\begin{tabular}{lrr}
\toprule
Label     & HealthRelease & HealthStory \\
\midrule
Fake      &       198      &   374    \\ 
Non\_Fake &      231         &  1036    \\ 
\midrule
Total News     &    429        &   1410         \\
\bottomrule
\end{tabular}
\end{center}
\caption{FakeHealth repository: Number of news along with their labels in the collected datasets} 
\label{tab:fakehealth_labels}
\end{table}

\noindent
\textbf{Step 1 - Unshortening URLs:} The first step involves expanding URLs created from URL shortening services, such as bitly~\cite{bitly}, tinyurl~\cite{tinyurl}, etc. This step is required to identify different short URLs that correspond to the same expanded URL. This step increases the probability of having a URL match with pre-existing annotated URLs from other research projects related to the 2016 US elections, as we explain in the next step. During this step, we use the unshrtn\footnote{\url{https://github.com/DocNow/unshrtn}} tool to expand the URLs.\\
\textbf{Step 2 - Pre-existing Labels:} Our dataset includes webpages related to the 2016 US elections, a well-studied dataset with many related research projects available in open-source version control systems, such as GitHub and GitLab. Thus, during this step, we aggregate more than 25 related projects with annotated URLs and combine them into a single database with $\approx$0.5M labeled URLs. Note that we only focused on the ``fake'' and ``non-fake'' labels and exclude any other labeling schemes (i.e., humor, satire, etc.) present in the other open-source projects. \\
\textbf{Step 3 - Labeling Consistency check:} During this step we examine the labeling consistency across all the datasets we combine during Step 2. We have kept only consistent labels, while leaving the inconsistent ones for manual validation, as we will explain in the next step. \\
\textbf{Step 4 - Matching URLs and Labels:} During this step, we perform a URL match between the URLs that we extract from our dataset and the one we created by aggregating labels from other open datasets related to the 2016 US elections~\cite{BuzzFeed_1, FakeNewsC, FakeNewsC2}. 

The output of this step is depicted in Table~\ref{tab:labellingstats} (second column) \textit{``Initial Labeling''}. \\
\textbf{Step 5 - Manual Labeling / FactCheck:} To increase the total number of labeled URLs, we then turn our attention to the ``Unknown'' URLs of the Initial Labeling phase. We use four different FactChecking tools, 1. PolitiFacts~\cite{PolitiFacts}, 2. Media Bias/Fact Check (MBFC)~\cite{MBFC}, 3. FactCheck~\cite{FactCheck}, 4. Snopes~\cite{Snopes}, and we manually annotate a subset of the unknown URLs. Since this step is very time-consuming, we only focused on expanded URLs that correspond to more than one short URL (see Step 1 above) in our dataset. 

The final number of each label is depicted in Table~\ref{tab:labellingstats} (third column - \textit{``Manual Labeling''}), while the final column (\textit{``Difference''}) depicts the difference between the Initial Labeling and the Manual Labeling steps. \\

\subsection{FakeHealth datasets}
In order to evaluate our method we utilize two datasets of the FakeHealth repository~\cite{FakeHealth}.
Due to the twitter policy of protecting user privacy, the full content of user engagement and network are not allowed to be published by the authors, instead, the authors provide a useful API available at \url{https://github.com/EnyanDai/FakeHealth}. The API provides the code and details on how to download the full content of users social engagements and network. Using the provided API we collect all related information, a summary is depicted in Tables~\ref{tab:fakehealth_data} and \ref{tab:fakehealth_labels}. Note that the final numbers reported in the above tables is lower by $\approx1.1\%$ than the numbers reported by the original authors. This can be attributed (1) to changes on behalf of the twitter users that choose to disallow public access of there tweets, (2) the tweeter itself delete the tweet due to some internal policies, (3) or the tweeter user account has been deleted. 

\section{The Meta-Graph Approach}
\label{metagraph}
Before we give the meta-graph's construction details, we first provide a formal definition of the data structure.

\begin{definition}[Meta-graph]
\label{def:meta}
Let $\mathcal{G} = (V, E)$ be a graph with vertex set $V$ and edge set $E$. Let also $X_{v_i}$ and $R_{e_{i,j}}$ be the feature vectors of node $v_i$ and of edge $e_{i,j}$ respectively. We call $\mathcal{G}$ a \textit{meta-graph} constructed from a set of events in a social network, if each event corresponds to a vertex $v_i \in V$, and the feature vectors $X_{v}$ and $R_{e_{i,j}}$ encode both user and event relational information.
\end{definition}

The node feature vectors in Definition \ref{def:meta} encode three types of features: user attributes, tweet content, and cascade structural information. A minimalist node feature vector $X_{v}$ can be described as
\begin{equation}
    X_{v} = (C_{\text{emb}} || X_u || T_{\text{emb}} || S ...), 
\label{eq:node-features}
\end{equation}
where "$||$" is the vector concatenation operation, $C_{\text{emb}}$ is the cascade embedding vector (obtained by using some graph embedding method, in our case DeepWalk \cite{DeepWalk}), $X_u$ are the concatenated feature vectors of all users present in the cascade, and $T_{\text{emb}}$ is the text embedding vector \cite{bert}, provided that the tweet also has the text content. $S$ is a vector of sentiment analysis scores if one such vector can be constructed, depending on the presence of text within the cascade tweets. The user feature vectors concatenated into $X_u$ contain information related to the user account itself, such as date of creation, number of followers, number of tweets, geolocation data, topic categories, sentiment analysis label and scores, etc. This formulation allows trivial expansion of node features by concatenating representative vectors from other sources when available.

The meta-graph's edges are initially constructed based on common users or a common topic (URL or tweet text). The corresponding edge feature attributes encode cascade similarity and tweet content similarity, once more defined in a very general manner. An example of a cascade feature vector is:
\begin{equation}
    R_{e_{i,j}} = (N_{u_{i,j}} || V_{i,j} || H_{i,j} ...), 
\label{eq:edge-features}
\end{equation}
, where $N_{u_{i,j}}$ represents a one-element vector containing the number of common users in cascades $i$ and $j$, $V_{i,j}$ is the value of a graph similarity metric \cite{graph_sim1, graph_sim2, graph_sim3, graph_sim4} applied to the two cascades. $H_{i,j}$ stands for a content similarity metric \cite{text_sim} between the contents of the original root tweets (and possibly retweets) of the two cascades, if available. The edge feature vector can, similar to the node feature vector, be trivially expanded to include additional relational information between cascades. 

\begin{figure*}[ht]
  \centering
  \includegraphics[width=0.95\linewidth]{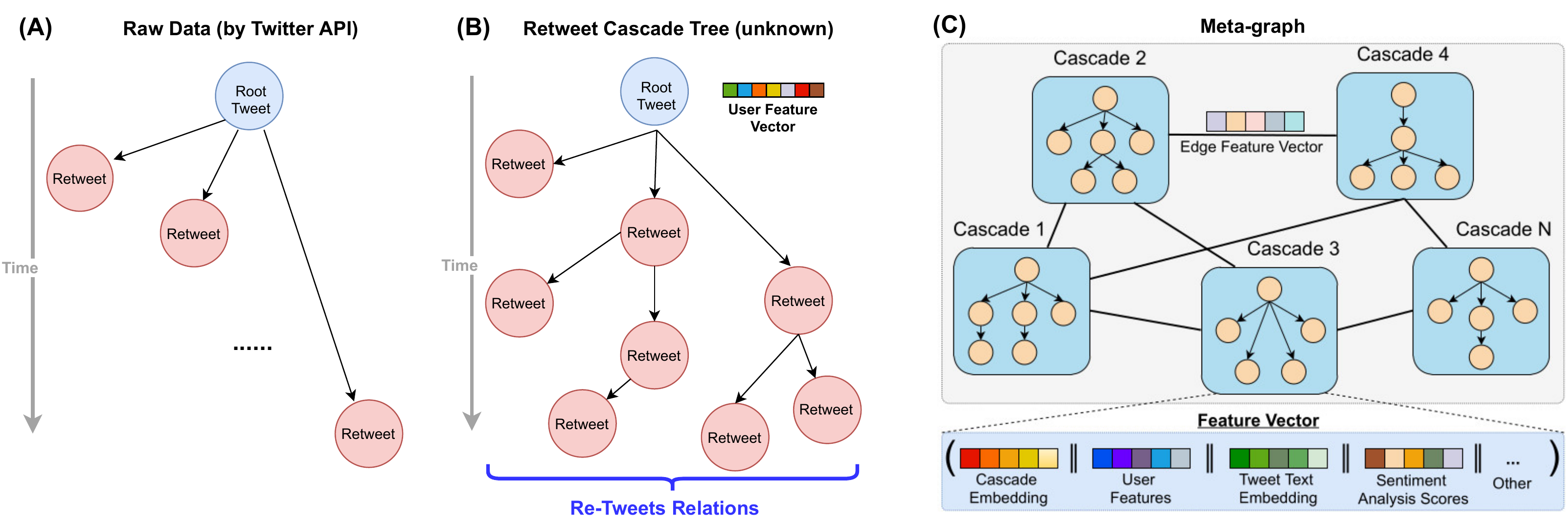}
	\caption{(a) The raw data that are provided by Twitter API correspond to a star graph structure. (b) The true retweet cascade tree, which is actually unknown, highlights the root tweet and subsequent retweets i.e the true series of retweet events. (c) The proposed meta-graph data structure: every node represents a Twitter cascade. The cascade features are given by the cascade vector embedding; the user features vector; the tweet text embedding and sentiment analysis scores (if retweet text is available). The edge features represent cascade structural similarity, i.e number of common users. This can easily be expanded with additional attributes, either from current or external datasets.}
	\label{fig:metagraph_all}
\end{figure*}

\subsection{The Meta-Graph Construction}
\label{submetagraph_construction}
There are two main obstacles to be addressed in order to apply our methodology to the data provided by Twitter. First, the actual social network -- the follower graph -- is mostly unknown; the users' follower lists are not always accessible. Second, the raw data returned by the Twitter API have limited information (by design) regarding the source of influence in a given retweet cascades. The only available information provided is the root--tweet and the root--user for a given retweet. In other words, all the retweeters have been influenced by the root--user. This star--like cascade structure (Figure~\ref{fig:metagraph_all}(a)) does not depict the true chain of retweet events, which in fact is a \textit{tree}, like the one presented in Figure~\ref{fig:metagraph_all}(b). 

We use the following steps to address the above issues:
\noindent
\textbf{Step 1:} Construct a graph that approximates the Twitter social network.\\
\textbf{Step 2:} Map a cascade to social affiliations of the users that participate in the cascade. Since the cascade is a subgraph of the social network we compute its embedding in a low dimensional space.\\
\textbf{Step 3:} Construct the final meta-graph.

We have to note that, we concentrate our analysis on pure retweet cascades, only. In this way, we ensure that the text that has been diffused by a given cascade (a chain of retweets) was exactly the same with the original/root tweet-text. For this reason, we have excluded the ``quotes'' from the meta-graph construction. A quoted-tweet is a special case of retweet, where the retweeter has added an additional text above the original one.

\subsubsection{The Social network}
We leverage the users' activity as it is recorded in the data to construct an approximation of the \text{follower graph} -- the true social network which is not publicly available to a large extent. In the Twitter platform, the interactions between users belong in three categories;
\textit{replies}, \textit{retweets}, and \textit{quotes} -- a special form of retweet -- and \textit{mentions}. Based on these actions, we construct a graph/network of interactions between the users. We map users to nodes and directed edges to interactions. For example, if a user--i has replied to a user tweet--j, then we add the edge $(i,j)$. The direction of the edge implies that $i$ is a \textit{follower} of $j$, while the reverse direction represents the information flow from $j$ to $i$. In conclusion, we map users to nodes and use the interactions between users to define the edges. This process outputs a multi--graph, where many edges may connect the same pair of users. For this reason, we discard the duplicate edges keeping only the earliest one.

\textbf{US Elections dataset:} The overall graph has 9.32M users/nodes connected with 84.1M directed edges. For the purpose of our analysis, the final social network is represented by the induced subgraph formed from the 3.63M users -- retweeters and root-users who participated in the retweet cascades -- who are connected with 61.05M directed edges.

Regarding the two FakeHealth datasets: In the HealthRelease we have 9,055 total users that participate in the retweet cascades. The corresponding social network counts 8,218 users (nodes) connected by 10,510 edges. In the HealthStory the total number of users is 64,593. The corresponding social network consists of 57,851 users and 88,750 edges.

The FakeHealth repository includes the \textit{user-following} adjacency lists who represent the ground-truth social networks. Specifically, in the HealthRelease we have 8,566 users connected by 177,866 edges. The HealthStory consists of 62,011 users and 3,402,241 edges. Having this information available, We compare the ``empirical'' social networks which we constructed based on the users' actions with the ground user-following relations that are available for the HealthRelease and HealthStory. In short, for both datasets, we constructed the “empirical” social network based on the actions between the users (mentions, replies, retweets). We restrict our attention only to those users who participated in retweet cascades, since only this graph region is involved in the meta-graph method. Then, we compared the ``empirical'' edges (i.e. relations) with the ground-truth ones. The 65\% and 59\% of the ``empirical'' edges for the HealthRelease and HealthStory respectively, appear in the ground-truth. Although these numbers are not very high they do not affect the overall validity of the meta-graph method. Our goal is not to predict the ground-truth social network but to use past interactions among the users in order to construct a small graph (per cascade) where we can compute the DeepWalk embedding.

\begin{figure}[t]
	\centering
    \includegraphics[width=0.90\linewidth]{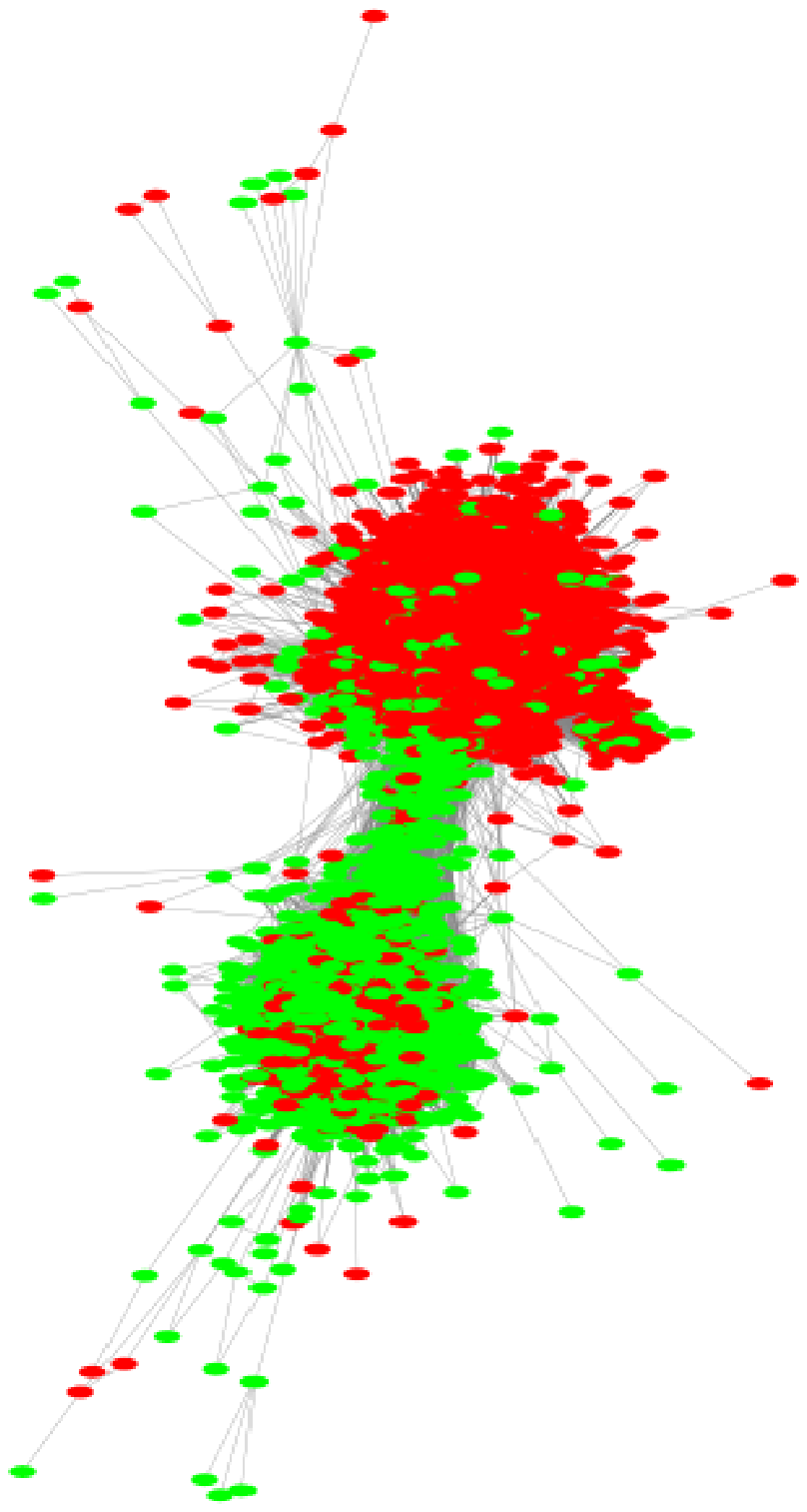}
	\caption{Meta-graph visualization for the US Elections dataset: Giant component of the meta-graph produced by the disparity filtering with $\alpha$=0.1. ``Fake'' nodes/cascades have been colored red, while the ``non--fake'' nodes are green. The meta-graph consists of 8291 cascades connected by 1,363,702 undirected edges and it has 3 connected components. The giant component counts 8282 nodes and 1,363,680 edges.}
	\label{fig:metagraph_image}
\end{figure}

\begin{figure*}[t]
	\centering
	\includegraphics[width=0.40\linewidth]{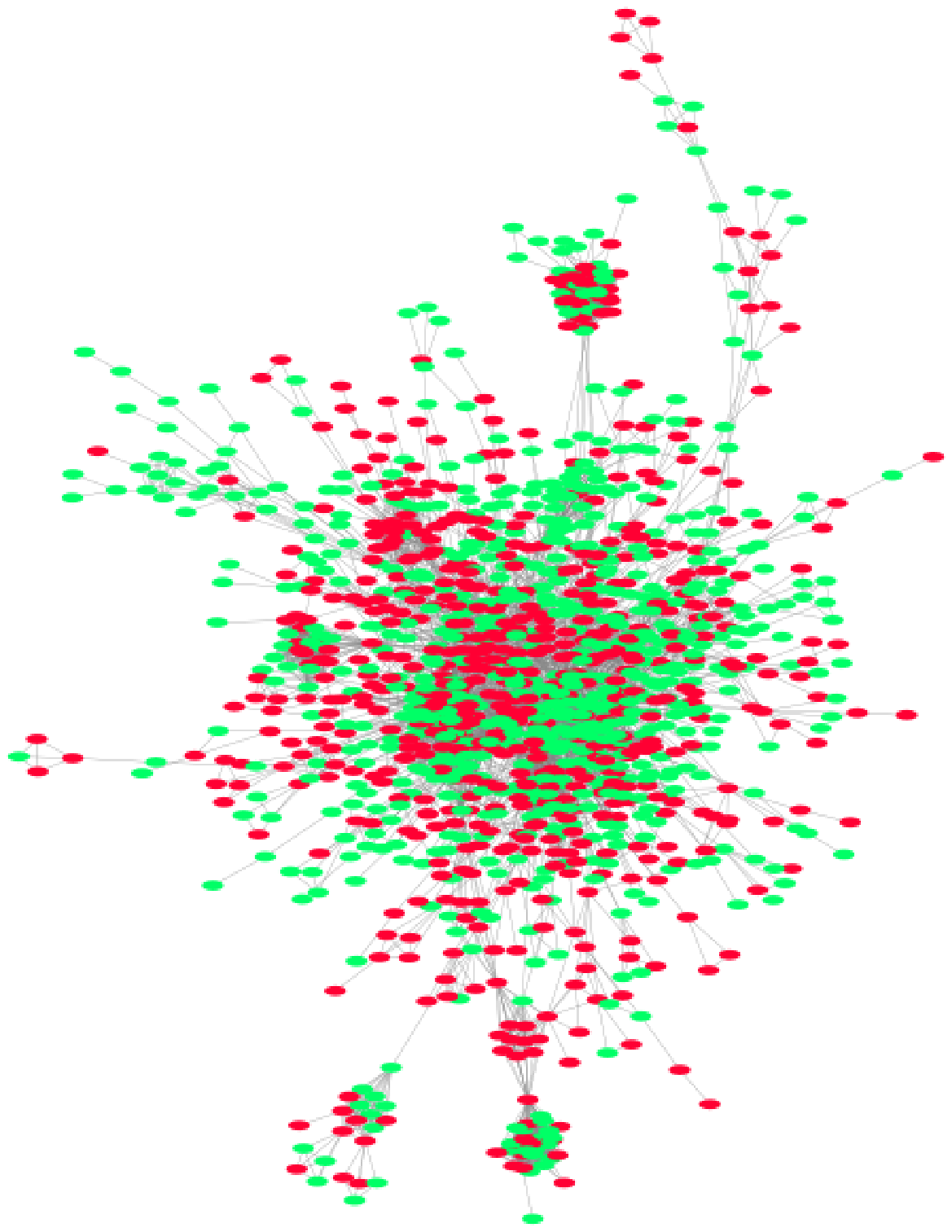}	
	\includegraphics[width=0.4\linewidth]{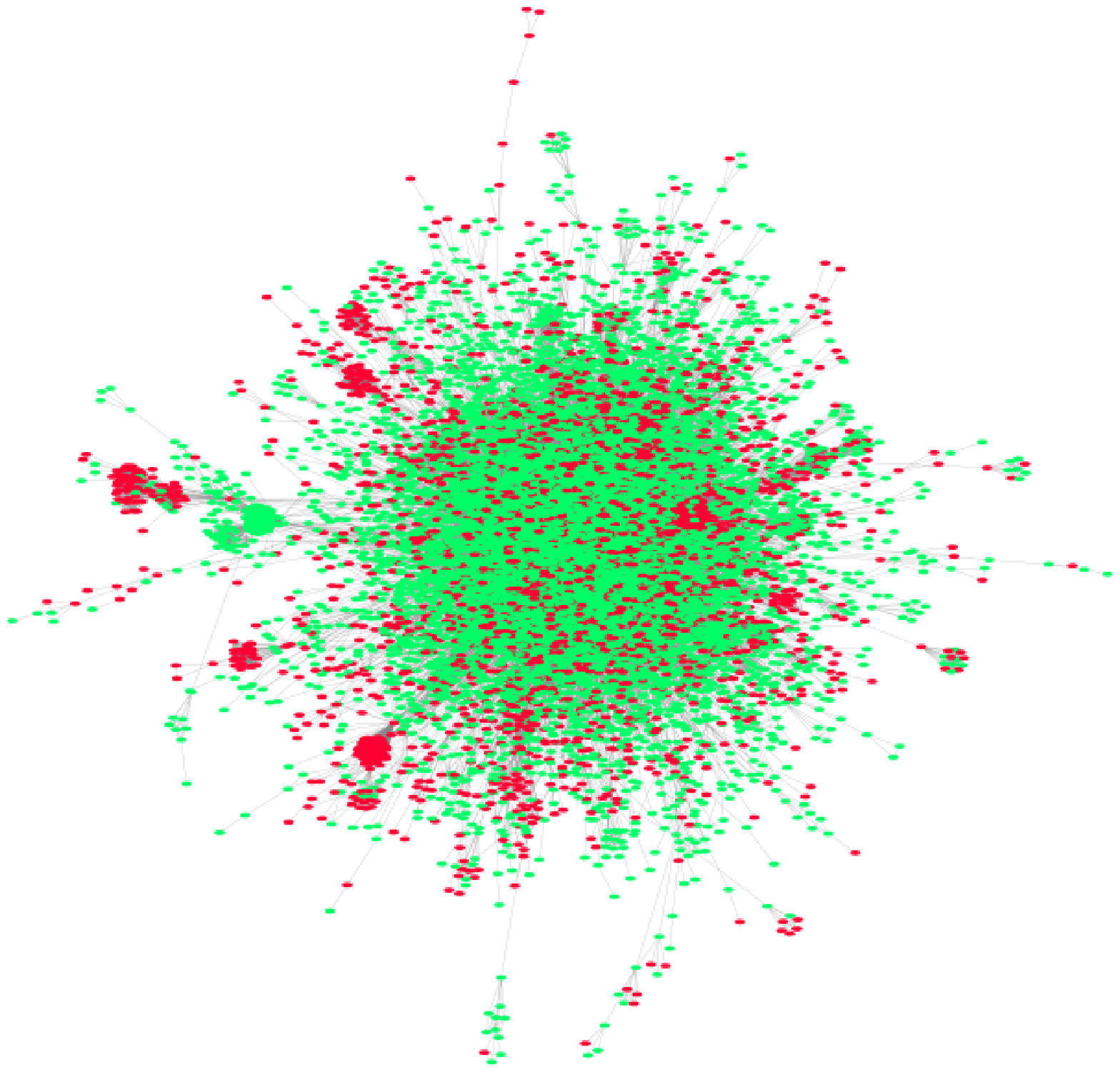}
	\caption{HealthRelease (left-hand figure) and HealthStory meta-graphs. ``Fake'' nodes/cascades have been colored red, while the ``non--fake'' nodes are green. Both are connected graphs.}
	\label{fig:metagraph_FakeHealth}
\end{figure*}

\subsubsection{From Cascades to Graph Embeddings}
\label{subsubsec:subgraphs}

As we mentioned previously, a retweet cascade is a series of chain events upon the same root--tweet. Some of the users have directly retweeted the root--tweet, whereas some others have retweeted a retweet of a friend on the same root--tweet (retweet of a retweet). This tree--like structure is the true retweet cascade (see Figure~\ref{fig:metagraph_all}(b)) and reflects the diffusion path of information that has been transmitted by the users in the social network. The problem we face here is that the data provided by Twitter do not represent the true cascades. The raw data contain only the retweet \& retweeter IDs and the root--tweet \& the root--user IDs. Hence, it is unknown who was influenced by whom during the retweeting process. The raw--data correspond to a star--like graph where all the retweeters are connected with the root--user. As a result, this form does not provide sufficient structural information. This is a well-known problem in the literature and many methods have been proposed to estimate the true diffusion path~\cite{Goel2015:virality}.

To address this problem, we leverage the social network we have constructed. We construct a subgraph formed by the interactions that the retweeters of this cascade have had in the past. Specifically, for a given retweeter $i$ who performed her retweet at date $t$, we identify which friends she had before the date $t$ and which of them belong to the set of retweeters of this cascade. Then we append this set of edges (if any) with the star--like structure, where each retweeter is connected with the root--user. Finally, we discard any duplicate edges and produce the undirected version. By this method, the subgraph is always connected. The extreme case occurs when the retweeters did not have any previous interaction. Then, the resulted graph coincides with the raw data collected from Twitter. In summary, the subgraph is just the star--like graph (i.e., the raw Twitter data) enhanced by the retweeters' social relations. This construction per cascade represents the social structure that the participants of the cascade had before being activated and be able to perform their retweet.

Each cascade should use the corresponding subgraph as a feature that will reflect the structural (social) relation that the cascades' participants had. To achieve that, in a consistent way, we 
produce the embedding of each subgraph to a low dimensional space. We use the DeepWalk 
algorithm~\cite{DeepWalk} and specifically the \textit{Karate Club} extension library's implementation for NetworkX~\cite{karateclub}. The DeepWalk embedding is a $N \times 128$ matrix, where $N$ is the 
number of users that participated in the cascade. We applied the default parameters of this implementation, that is: (i) $\text{Number of random walks}=10$; (ii) $\text{Length of random walks}=80$; (iii) $\text{Dimensionality of embedding}=128$. 

\subsubsection{Topics and Sentiment Analysis}
\label{subsec:topics_sentiment}

While the previous set of features were concerned with user similarity and relational information extracted from the social network, the final features, which complete the meta-graph, are focused on semantics and subject-based grouping of tweet-retweet cascades. For this purpose and for the US Elections dataset only, we carry out two tasks, topic detection and sentiment analysis of the root-tweet content. For the former, as the initial number of topics covered by our dataset is unknown, we employ a Hierarchical Dirichlet Process (HDP) \cite{hdp1, hdp2}, using the Tomotopy library \cite{tomotopy}. Three parameters of the HDP model are changed from the default values, namely $\text{min\_cf} = 5$, $\text{gamma} = 1$ and $\text{alpha} = 0.1$, as these produce the best results. For term weighing, we produce three instances of the model with equal term weighing, Point-wise Mutual Information-based weighing and Inverse Frequency term weighing (low weights for terms with high occurrence and vice-versa). After training the three versions of the HDP model on our tweet content, we use them in order to assign a topic ID to each tweet-retweet cascade. 

The second step in the process is to perform sentiment analysis on the root-tweet of each cascade. This is done using a pre-trained BERT language model, with the help of the Transformers library \cite{transformers}. Along with the three topic identifiers obtained above, the sentiment label and sentiment score form the semantic features used in our approach.

The semantic information, together with the embeddings obtained in the previous section, represent the features of each individual cascade, in the case of graph-level classification, or of each individual node, in the meta-graph, for the node-classification task. In the case of cascades, it is assumed that retweets share the topic and sentiment of the root tweet. 

\subsubsection{The meta-graph}
The final step is the construction of the meta-graph itself. That is, a meta--structure represented by a graph that has the retweet cascades as nodes. The edges of the meta-graph are defined by the proximity among the cascades in terms of their retweeter population. 

In particular, two cascades $i$ and $j$ are connected by an edge if $ |RT_{i} \cap RT_{j}| \geq 1$, where $RT_{i}$ and $RT_{j}$ are the set of retweeters of $i$ and $j$. In other words, two cascades are connected by an edge when they share at least one retweeter. In conclusion, the meta-graph is always an undirected weighted graph, where the weight of each edge is equal to the number of retweeters that participate in both nodes/cascades. In case that the meta-graph is very dense, we apply the disparity filtering method for undirected weighted graphs~\cite{disparity_filter}. Given a significance level $\alpha$, the disparity filter identifies which edges should be preserved in the graph (see Equation-2 in~\cite{disparity_filter}).

US Elections dataset: The meta-graph of 8323 labeled cascades counts 8323 nodes/cascades connected by 15,946,910 undirected edges. Since the graph is extremely dense, we apply the disparity filter for several $\alpha$ values. In this way, we produce one filtered meta-graph for each $\alpha$ value. Then, we perform the classification in each meta-graph independently. We report the results in Table~\ref{tab:alpha_selection}. As an example, Figure~\ref{fig:metagraph_image} presents the giant component of the meta-graph that has been produced by the disparity filter for $\alpha = 0.1$. 8291 nodes/cascades out of the 8323 are connected by 1,363,702 undirected edges. The graph has 3 connected components. The giant component counts 8282 nodes and 1,363,680 edges.

FakeHealth datasets: In the HealthRelease the full version of the meta-graph counts 1969 cascades connected by 11,581 undirected edges (Figure~\ref{fig:metagraph_FakeHealth}, left-hand figure). In the HealthStory the full meta-graph consists of 13,263 cascades connected by 100,001 edges (Figure~\ref{fig:metagraph_FakeHealth}, right-hand figure). Since these graphs are not very large, we use the full version for the classification.


\begin{table}[]
\begin{center}
\begin{tabular}{cccc}
\hline
$\alpha$ & \begin{tabular}[c]{@{}c@{}}Number \\ of \\ Nodes\end{tabular} & \begin{tabular}[c]{@{}c@{}}Number \\ of\\  Edges\end{tabular} & \begin{tabular}[c]{@{}c@{}}Accuracy \\ of \\ GCNConv\\ (\%)\end{tabular} \\ \hline
0.01 & 8171 & 300,330   & 83.96 \\
0.02 & 8221 & 434,643   & 84.06 \\
0.03 & 8246 & 555,934   & 84.20 \\
0.04 & 8260 & 673,595   & 84.46 \\
0.05 & 8262 & 788,212   & 84.44 \\
0.06 & 8268 & 902,797   & 84.56 \\
0.07 & 8277 & 1,016,344 & 84.68 \\
0.08 & 8285 & 1,131,124 & 85.04 \\
0.09 & 8290 & 1,247,597 & 85.17 \\
0.1 & 8291 & 1,363,702  & 85.34 \\
0.2 & 8310 & 2,606,347  & 86.52 \\
0.3 & 8311 & 4,079,032  & 86.60 \\
0.4 & 8315 & 5,742,849  & 87.64 \\
\textbf{0.5} & \textbf{8319} & \textbf{7,763,018} & \textbf{87.70} \\
0.6 & 8319 & 9,899,401  & 87.66 \\
0.7 & 8319 & 12,536,442 & 86.91 \\
0.8 & 8320 & 14,866,835 & 87.12 \\
0.9 & 8320 & 15,829,764 & 86.45 \\ \hline
\end{tabular}
\caption{$\alpha$ parameter selection using GCNConv as the best performing model.}
\label{tab:alpha_selection}
\end{center}
\end{table}


\begin{table}[]
\begin{small}
\begin{tabular}{ccccc}
\hline
Model                                                                              & \multicolumn{2}{c}{\begin{tabular}[c]{@{}c@{}}Cascade\\  Classification\end{tabular}}                         & \multicolumn{2}{c}{\begin{tabular}[c]{@{}c@{}}Metagraph \\ Node Classification\end{tabular}}                   \\ \hline
\multicolumn{5}{c}{2016 US Presidential Elections Dataset}                                                                                                                                                                                                                                                          \\ \hline
\begin{tabular}[c]{@{}c@{}}Without Topics\\ and Sentiment \\ Analysis\end{tabular} & \begin{tabular}[c]{@{}c@{}}Accuracy\\ (\%)\end{tabular} & \begin{tabular}[c]{@{}c@{}}F1 \\ Score\end{tabular} & \begin{tabular}[c]{@{}c@{}}Accuracy\\  (\%)\end{tabular} & \begin{tabular}[c]{@{}c@{}}F1 \\ Score\end{tabular} \\ \hline
GCNConv                                                                            & \textbf{83.30}                                          & \textbf{0.721}                                      & \textbf{86.75}                                           & \textbf{0.812}                                      \\
GATConv                                                                            & 83.28                                                   & 0.716                                               & 85.52                                                    & 0.793                                               \\
HypergraphConv                                                                     & 82.17                                                   & 0.693                                               & 85.67                                                    & 0.791                                               \\
SAGEConv                                                                           & 82.23                                                   & 0.702                                               & 85.12                                                    & 0.787                                               \\ \hline
\begin{tabular}[c]{@{}c@{}}With Topics\\ and Sentiment\\  Analysis\end{tabular}    & \begin{tabular}[c]{@{}c@{}}Accuracy\\ (\%)\end{tabular} & \begin{tabular}[c]{@{}c@{}}F1 \\ Score\end{tabular} & \begin{tabular}[c]{@{}c@{}}Accuracy\\  (\%)\end{tabular} & \begin{tabular}[c]{@{}c@{}}F1 \\ Score\end{tabular} \\ \hline
GCNConv                                                                            & \textbf{84.23}                                          & \textbf{0.748}                                      & \textbf{87.70}                                           & \textbf{0.830}                                      \\
GATConv                                                                            & 83.74                                                   & 0.732                                               & 86.62                                                    & 0.822                                               \\
HypergraphConv                                                                     & 83.91                                                   & 0.731                                               & 86.67                                                    & 0.825                                               \\
SAGEConv                                                                           & 83.42                                                   & 0.727                                               & 86.84                                                    & 0.810                                               \\ \hline
\multicolumn{5}{c}{Health Release Dataset}                                                                                                                                                                                                                                                                          \\ \hline
\begin{tabular}[c]{@{}c@{}}With Topics \\ and Sentiment\\  Analysis\end{tabular}   & \begin{tabular}[c]{@{}c@{}}Accuracy\\ (\%)\end{tabular} & \begin{tabular}[c]{@{}c@{}}F1 \\ Score\end{tabular} & \begin{tabular}[c]{@{}c@{}}Accuracy\\  (\%)\end{tabular} & \begin{tabular}[c]{@{}c@{}}F1 \\ Score\end{tabular} \\ \hline
GCNConv                                                                            & \textbf{86.54}                                          & \textbf{0.890}                                      & \textbf{88.07}                                           & \textbf{0.936}                                      \\
GATConv                                                                            & 84.43                                                   & 0.753                                               & 87.81                                                    & 0.903                                               \\
HypergraphConv                                                                     & 85.76                                                   & 0.794                                               & 86.60                                                    & 0.892                                               \\
SAGEConv                                                                           & 84.82                                                   & 0.758                                               & 86.74                                                    & 0.896                                               \\ \hline
\multicolumn{5}{c}{Health Story Dataset}                                                                                                                                                                                                                                                                            \\ \hline
\begin{tabular}[c]{@{}c@{}}With Topics \\ and Sentiment\\  Analysis\end{tabular}   & \begin{tabular}[c]{@{}c@{}}Accuracy\\ (\%)\end{tabular} & \begin{tabular}[c]{@{}c@{}}F1 \\ Score\end{tabular} & \begin{tabular}[c]{@{}c@{}}Accuracy\\  (\%)\end{tabular} & \begin{tabular}[c]{@{}c@{}}F1 \\ Score\end{tabular} \\ \hline
GCNConv                                                                            & 59.23                                                   & 0.616                                               & 60.16                                                    & 0.751                                               \\
GATConv                                                                            & \textbf{60.18}                                          & \textbf{0.632}                                      & \textbf{61.03}                                           & \textbf{0.757}                                      \\
HypergraphConv                                                                     & 56.80                                                   & 0.587                                               & 58.70                                                    & 0.735                                               \\
SAGEConv                                                                           & 57.36                                                   & 0.590                                               & 57.36                                                    & 0.726                                               \\ \hline
\end{tabular}
\end{small}
\caption{Classification results obtained with four different graph neural network types, on three datasets: the 2016 Presidential Election dataset and the FakeHealth package, composed of the Health Release and Health Story datasets.}
\label{tab:results}
\end{table}

\textbf{Node features:} The features of a given node/cascade with $N$ users (the retweeters plus the root--user) are the following: (1) The DeepWalk embedding of the correspondent subgraph -- a $N \times 128$ matrix; (2) The dates of users' account creation; (3) The maximum value of users' followers count; (4) The maximum value of users' friends count; (5) The maximum value of users' statuses count; (6) The maximum value of users' favorites count; (7) A Boolean identifier whether the users' account is verified; (8) The language of the users' account. Note that features 2 to 8 comprise an array of length $N$. The maximum values per user are based on the retweets that the user has posted; (9) The most representative topic of the root--tweet based on the three topic models -- three values. Regarding the two FakeHealth datasets, we set as topics the news' ``tags'' and ``category''. This information is provided by the ``reviews'' in the FakeHealth repository; (10) The sentiment \textit{(label, score)} of the root--tweet -- two values, either $(2, score)$ for 'Positive' label or $(1, score)$ 'Negative' label.

Regarding the US Elections dataset, we note that we analyse the retweet cascades in which at least one URL is embedded in the root--tweet. Since the labeled part of the data is the URLs, we use these labels as ground--truth. We focus our attention on the 6525 URLs that have been labeled as ``fake'' and ``non--fake'' (Table~\ref{tab:labellingstats}). For each URL, we collect the cascades that had this URL embedded in their root-tweets. Moreover, there is no one--to--one correspondence between the cascades and the URLs -- multiple cascades may have spread the same URL. This is why the number of labeled cascades is larger than the number of labeled URLs. Moreover, a cascade might contain several URLs, ``fake'' and ``non--fake'' ones. In this case, we discard these cascades.

\section{Evaluation of Performance Benefits}
\label{experiments}

In order to evaluate the benefits of the meta-graph method, we perform a series of experiments aiming to address the following research questions: How effective is the meta-graph structure for the cascade classification?
Will the topics (broad subject shared by a set of tweets) and sentiment labels (positive or negative sentiment, together with the confidence score) improve the separation of classes if they are included in the features? 

\begin{figure*}[t]
	\centering
	\includegraphics[trim={3cm 2.5cm 0 3cm}, width=1.08\textwidth]{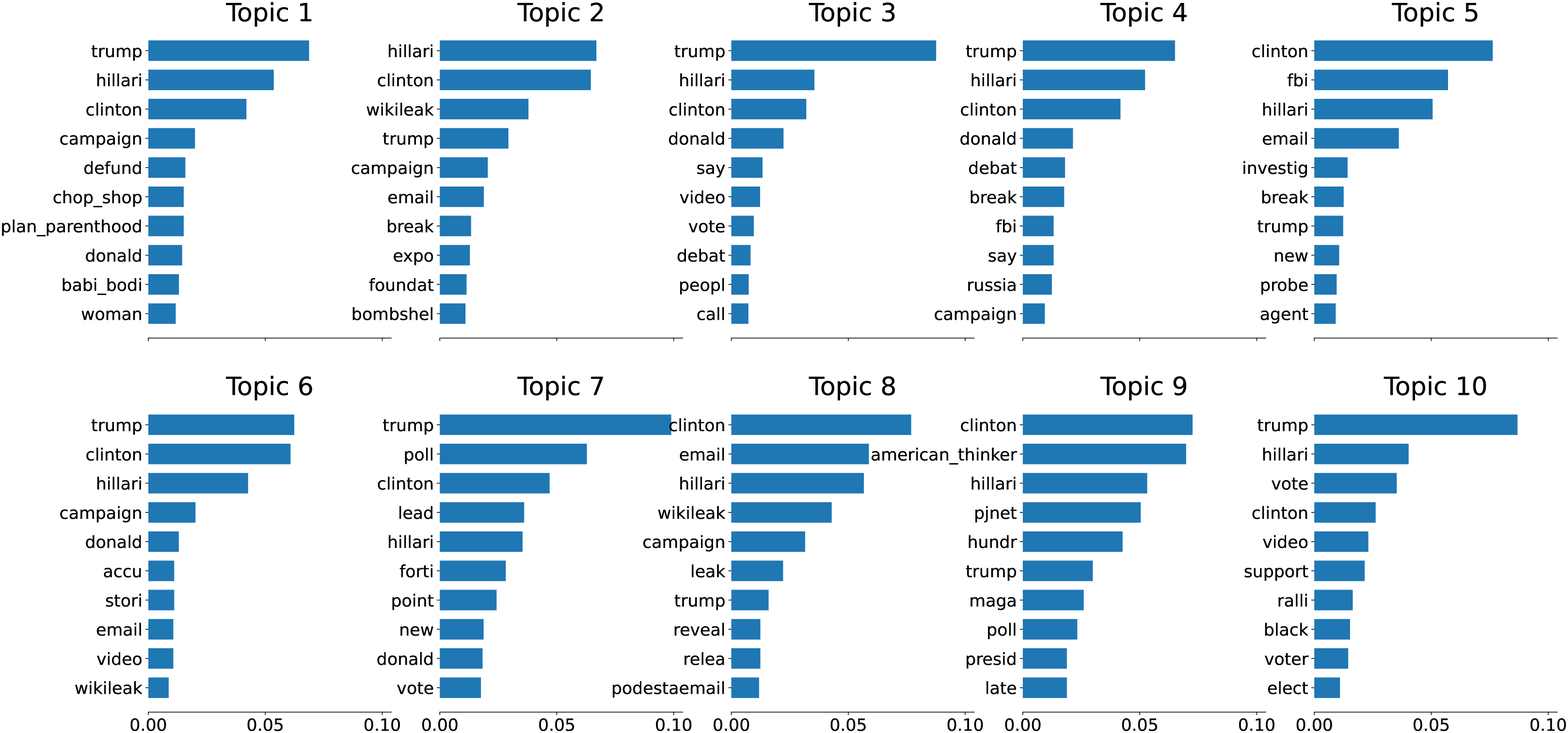}

    \small
    \resizebox{\textwidth}{!}{%
    \begin{tabular}{p{0.05cm}p{9.18cm}p{14.21cm}} 
        \toprule
        \# & News Article Title & Source \\ 
        \hline
        1     & Clinton: Planned Parenthood videos 'disturbing' & \url{https://edition.cnn.com/2015/07/29/politics/hillary-clinton-planned-parenthood-anti-abortion/index.html} \\ 
        2     & Hillary Clinton Email Archive - Wikileaks & \url{https://wikileaks.org/clinton-emails/} \\
        3     & 4 key moments from tonight's messy debate & \url{https://edition.cnn.com/politics/live-news/presidential-debate-coverage-fact-check-09-29-20/index.html} \\
        4     & Transcript of the Second Debate & \url{https://www.nytimes.com/2016/10/10/us/politics/transcript-second-debate.html} \\
        5     & FBI probing new emails related to Clinton case & \url{https://www.cnbc.com/2016/10/28/fbi-probing-new-clinton-emails.html} \\
        6     & Emails Related to Clinton Case Found in Anthony Weiner Investigation & \url{https://www.nbcnews.com/news/us-news/fbi-re-open-investigation-clinton-email-server-n674631} \\
        7     & Hillary Clinton Leads Donald Trump by 14 Points Nationally in New Poll & \url{https://time.com/4546942/hillary-clinton-donald-trump-lead-poll/} \\
        8     & John Podesta email article & \url{https://en.wikipedia.org/wiki/Podesta\_emails} \\
        9     & Are Fake News Polls Hiding a Potential Trump Landslide? & \url{https://www.americanthinker.com/articles/2020/07/are_fake_news_polls_hiding_a_potential_trump_landslide.html} \\
        10    & Russian trolls' chief target was 'black US voters' in 2016 & \url{https://www.bbc.com/news/technology-49987657} \\
        \bottomrule
        \end{tabular}
    }
	\caption{US Elections dataset: \textbf{Top:} The top-10 Topics by the Hierarchical Dirichlet Process (HDP) when all terms weighted equally. \textbf{Bottom:} The corresponding web-articles of each topic.}
	\label{fig:top10_topics}
\end{figure*} 

In order to address these questions, we follow two  classification strategies: (1) First, we ignore the information that is represented by the meta-graph edges and we reduce the problem to a graph classification task. Each retweet cascade is represented by a sub-graph of the social network (see Subsection \ref{submetagraph_construction}), namely a small graph which consists of those users who have been involved in the cascade. The users' features are those we presented in Subsection~\ref{submetagraph_construction}. Moreover, we assume that each user inherits the topic and sentiment label from the root--tweet. In other words, for a given cascade, all users have the same topic and sentiment label as feature. (2) The second classification approach is node-classification in the meta-graph level, where each node corresponds to a cascade (see Figure~\ref{fig:metagraph_all}(c)).

In other words, we classify the nodes of the meta-graph based on the meta-graph structure together with the nodes/cascades features. Now the nodes correspond to individual cascades. In both approaches, we rely on the same sub-graphs from the constructed meta-graph. In addition to these and in order to justify the need for topic modeling and sentiment analysis, we compare the two classification strategies with and without the use of topics and sentiments as nodes features. The intuition is that the topic modeling and sentiment analysis introduce semantic information in the meta-graph structure, providing a quantifiable measure of agreement as well as topical relational information to other nodes.

Our findings show that: (1) This method of connecting cascades/nodes is effective enough to improve by 3\%--4\% the classification accuracy. In the meta-graph two cascades are connected by an edge if they have more than 10 retweeters in common. (2) When we include the topics and sentiment labels as features, the accuracy of all classifiers is increased by approximately 1\%.

Regarding the actual methods, we applied the following state-of-the-art graph neural networks (GNNs) for both graph and node classification: \textit{GCNConv}~\cite{gcn}, \textit{SAGEConv}~\cite{sage}, \textit{HypergraphConv} ~\cite{hypergraphconv} and \textit{GATConv} ~\cite{gat}.

For the graph classification task, we have applied the algorithms to the 8318 sub-graphs (after filtering contradictory labels). As the treated graphs are relatively small, and GNNs generally do not benefit from an increase in the number of layers \cite{gnn_depth1, gnn_depth2, gnn_depth3, gnn_depth4}, we restrict the models to one graph feature extraction layer and one dense layer. We also use a 60-20-20 train-validation-test split. The GNNs have been implemented in PyTorch Geometric \cite{pyg}, maintaining the following hyperparameters constant across experiments:

$$\text{\# of epochs} = 50, \text{dropout}=0.8, \text{learning rate} = 10^{-4}$$

In Table \ref{tab:results} we present the obtained results, which show a 3\%-4\% gain in accuracy in favor of the meta-graph method, i.e. for the node classification task. We get the best performance when we include the topics and sentiment labels. For state-of-the-art algorithms this represents a significant gain, and primarily depends on the inclusion of relational information between retweet cascades. Regarding the topic modeling and sentiment analysis, we observe that the 1\% gain in accuracy, is nonetheless an indicator that the discovered topics and sentiment labels indeed lead to an improvement to the separation of the classes. 

Due to the imbalance of the dataset, we also present F1 scores for all our experiments. We observe a general consistency in terms of the advantage provided by the meta-graph method compared to regular cascade classification. In particular, the extra relational information present in the meta-graph does contribute to the reduction of the number of false positives and false negatives. 

In Table \ref{tab:results} we also present results for two health-related misinformation datasets, namely HealthRelease and HealthStory. These smaller datasets have the added disadvantage of simpler cascade structures. While most cascades present contain a small number of users (sometimes just a tweet-retweet pair), larger cascades tend to have a star-graph shape. This reduces the structural information that our proposed method uses, with the help of graph embedding algorithms, such as DeepWalk.

We conclude the analysis by portraying a representative sample of the topic modeling in order to emphasize the effectiveness of this approach. Figure~\ref{fig:top10_topics} depicts the top--10 topics based on the HDP-model (Hierarchical Dirichlet Process) when all terms weighted equally. For each topic in the top--10 we plot the top--10 terms by their log-likelihood values. In addition, in Figure~\ref{fig:top10_topics} (Bottom),

for each top--10 topic we show a representative news article related to the period of the 2016 Presidential elections. For instance, the top terms of Topic-1 and Topic-2 reflect the "Planned Parenthood" debate and the \textit{Wikileaks} source related to Hillary's Clinton leaked emails, respectively.

\section{Discussion}
\label{discussion}

Our proposed method addresses the detection of disinformation in social networks, by exploiting network structure as well as post content and user sentiment. The example datasets focus on the Twitter social network, whose characteristic is given by the shortness of exchanged user messages. This has an effect on two components of our method: the sentiment analysis component and the text embedding. As some tweets may not contain any original content, and represent just links (retweets) to other user's posts, we have designed our method to be robust to this particular type of information scarcity, by exploiting network structure wherever possible. 

Network structure is an important aspect with a clear impact on classification results. The meta-graph is constructed from individual tweet-retweet cascades. As such, the structural information of the cascade is contained within a part of the corresponding features of the meta-graph node. These features are obtained by using graph embedding algorithms, for example DeepWalk, in order to capture cascade structure and connectivity patterns. In the case of very small cascades, or when the majority of cascades have star-graph shapes, the embedding vectors are very similar. The effects of this on the performance of our method are visible in Table \ref{tab:results}, for the specific case of HealthStory.

In the present paper we adopt a simple method for transferring the URL labels, namely the labels of the root tweet content, to the entire cascade, by assuming that the label of the cascade is given by the label of the content of its root tweet. This can, in some situations, create problems, for instance if the root tweet and all following retweets share a URL containing false information, with the explicit purpose of exposing it, the entire cascade will be labeled as disinformation. We deal with this potential issue by removing "quote tweets", containing additional text. If retweet contents are available, then one can easily expand this trivial labeling to take the already computed sentiment analysis scores into account. If the sentiment scores are strongly negative, a disagreement between the tweets and URL content can be detected and the label of the cascade can be adjusted accordingly.

Due to the nature and structure of our considered datasets, a direct benchmark comparison with similar algorithms such as FANG \cite{fang} and Hierarchical Propagation Networks \cite{hierarchical} is not possible, as the mentioned algorithms require the construction of different types of graphs, for which the considered datasets do not contain the relevant information. An example is the heterogeneous graph required for the FANG algorithm, where both articles/posts and users are interconnected nodes.

The strength of our method relies on it's capacity to treat information-scarce datasets as long as structure can be exploited, due to the graph embedding features playing a central role. This also displays the inadequacy of the method to datasets which have very limited graph structure, such as the HealthStory dataset, on which the performance both in terms of accuracy and F1 score is very low. In limit-cases, for example in treating isolated nodes, the node classification task naturally reduces to a simple graph (cascade) classification task. As such, the meta-graph approach does not need to remove or provide special treatment to isolated nodes.

\section{Conclusion}
\label{conclusion}

Recent false information detection and classification methodologies rely on user features extracted from the social network, network structure, or the posting (in our case, tweet) content itself. This paper aims to unify these approaches via the use of a single data structure, which we call a \textit{meta-graph}. The meta-graph node features represent retweet cascades, containing information about the tweet-retweet event and the individual users taking part in it. Encoded within the node features, we also add the tweet content, where available. At the same time, the edge features contain feature vectors whose elements are similarity metrics between cascades. This information is beneficial when only a small number of cascades in the meta-graph are labeled.

By combining all available information about a social-network event into a single data structure, we provide a graph-specific classification algorithm with an informational-rich data format that allows it to outperform, in terms of classification accuracy values, approaches based on isolated elements (such as individual graphs). The additional similarity information among pairs of cascades is beneficial in semi-supervised classification settings because labels are routinely hard to obtain, and only a small fraction of the data may have them. 

Another advantage of the method is the size of the dataset itself. Even if the meta-graph contains considerably more information than the individual constituent cascades, the storage cost is not prohibitive. In our case, for 8008 nodes and 1,979,031 undirected edges, the dataset (either in meta-graph or cascade sub-graph form) occupies approximately 3 GB.

The presented formalism opens the door for a wide range of future extensions. The meta-graph can be generalized to include not just bipartite relations between events (cascades), but also multipartite ones, thus converting the data structure into a hypergraph. Another possible research direction is concerned with finding the optimal features to include for nodes and edges. This is far from trivial, as there is a large variety of parameters available, characterizing a tweet or a particular user, as well as many ways in which the graph structure of the cascades can be encoded. Finally, relation learning, similar to the case of knowledge graphs, can be considered, and learned edge features be added to existing ones. 

\section*{Acknowledgments}
This project has been funded by: (i) the European Union's Horizon 2020 Research and Innovation program under the Cybersecurity CONCORDIA project (Grant Agreement No. 830927) and under the Marie Skłodowska--Curie INCOGNITO project (Grant Agreement No. 824015); (ii) the TV-HGGs project (OPPORTUNITY/0916/ERC-CoG/0003), co-funded by the European Regional Development Fund and the Republic of Cyprus through the Research and Innovation Foundation.

\bibliography{disinformation_spread}

\end{document}